# Robust and Multifunctional Liquid-Metal Embedded Elastomers for Ultrastretchable Electronics: a Short Review

**Kaveh Alizadeh \***

**Abstract**

Soft electronics are a promising and revolutionary alternative for traditional electronics when safe physical interaction between machines and the human body is required. Among various materials architectures developed for producing soft and stretchable electronics, Liquid-Metal Embedded Elastomers (LMEEs), which contain Ga-based inclusions as a conductive phase, has drawn considerable attention in various emerging fields such as wearable computing and bio-inspired robotics. This is because LMEEs exhibit a unique combination of desirable mechanical, electrical, and thermal properties. For instance, these so-called multifunctional materials can undergo large deformations as high as 600% strain without losing their electrical conductivity. Moreover, the desperation of conductive liquid-metal inclusions within the entire medium of an elastomer makes it possible to fabricate autonomously self-healing circuits that maintain their electrical functionality after extreme mechanical damage induction. The electrically self-healing property is of great importance for further progress in autonomous soft robotics, where materials are subjected to various modes of mechanical damage such as tearing. In this short review, we review the fundamental characteristics of LMEEs, their advantages over other conductive composites, materials used in LMMEs, their preparation and activation process, and the fabrication process of self-healing circuits. Additionally, we will review the soft-lithography-enabled techniques for liquid-metal pattering.

\* Kaveh Alizadeh
Department of Mechanical Engineering
Amirkabir University of Technology, Tehran, Iran
Tel: +98 922 456 8586
E-mail: k_alizadeh@aut.ac.ir

## 1 Introduction

Merging the human body and smart machines and their intelligence offer promising and revolutionary solutions for the diagnosis and therapy of various diseases. While our body is composed of soft and deformable biological tissues – with mechanical stiffness from approximately 1 KPa to a few MPa, traditional microelectronics and machines are typically composed of intrinsically rigid and brittle materials that posse poor mechanical flexibility and stretchability **(Figure. 1)** [1-3]. Due to this extreme mechanical and physical mismatch between building components of conventional rigid electronic devices with soft biological tissues, traditional microelectronics and semiconductor technologies demonstrates poor capabilities for safe physical interaction with humans [1]. Thus, it is essential to exploit human-friendly and bio-compatible devices and interfaces to tackle this mechanical and physical mismatch when physical interaction between humans and machines is required.

Recent advances in materials have made it possible to design a broad range of soft materials – polymers, elastomers, and hydrogels – that can match biological tissue's mechanical properties and bridge the technological gap between machines and humans. Soft materials form a safe, long-term, and bio-compatible interface between hard devices and soft human skin and organs **(Figure. 2)** [3]. These materials with the elastic modulus in the order of kilopascals **(Figure. 1)** are so comfortable and stretchable that they can be attached to the skin without irritation or tissue damage.

However, In comparison to rigid metals – metals, ceramics, silicon, and plastics – soft materials typically have poor electrical and thermal properties [1]. By taking this fact into account, researchers have developed several strategies to integrate soft insulator materials with conductive electronic materials to develop multifunctional soft materials for stretchable electronics applications. These so-called multifunctional materials have the capability of undergoing large deformations under various mechanical loadings (e.g., tension, twisting, bending, etc.) without losing their electrical and thermal properties. The developed methods to design multifunctional soft materials for stretchable electronics include deterministic structures; or wavy 2D patterns on soft stretchable substrates [4, 5]; nanoparticle-filled polymers [6, 7], conductive fluid-based composites, and recently developed liquid-metal embedded elastomers (LMEEs), the focus of this manuscript [1, 2, 8, 9]. In these heterogeneous electronics, soft materials constitute the deformable host material. Wavy thin metallic films, nanomaterials such as carbon allotropes, conductive fluids, and liquid metals constitute the conductive part of stretchable electronics.

Among various methods and materials developed to fabricate multifunctional soft materials for stretchable electronics, LMEEs in which liquid metals are embedded within a continuous elastomer body at desired spatial locations promise revolutionary materials for next-generation wearable devices and soft robotics. Furthermore, these materials posse a unique property of ultrastretchability without sacrificing their electrical and thermal properties. For instance, by embedding Ga-based liquid metals within PDMS, reaching limit strains as high as 600% without losing electrical or thermal conductivity is quite possible [10]. Although in various applications like the field of wearable electronics, the applied strain may not exceed 10%, the ultimate aim of soft-matter technologies is to develop a general framework and technology that meets the extensibility requirements in a wide range of fields like soft sensors [11-13] and bio-inspired organism-like soft robotics [14-16]. Moreover, these multifunctional soft materials can also be designed to exhibit electrically self-healing properties under extreme mechanical damaging conditions.

This review focuses on liquid metal embedded elastomers, but we will first elaborate on other strategies and their merits and demerits over LMEEs. This manuscript's structure is as follows: the next section presents a brief review of the background of soft electronics, followed by the review of the fundamental characteristics of LMEES, materials employed for LMEEs, and their synthesis methods. Then, patterning techniques for LMEE-based soft electronics and the case study of self-healing LMEEs will be presented.

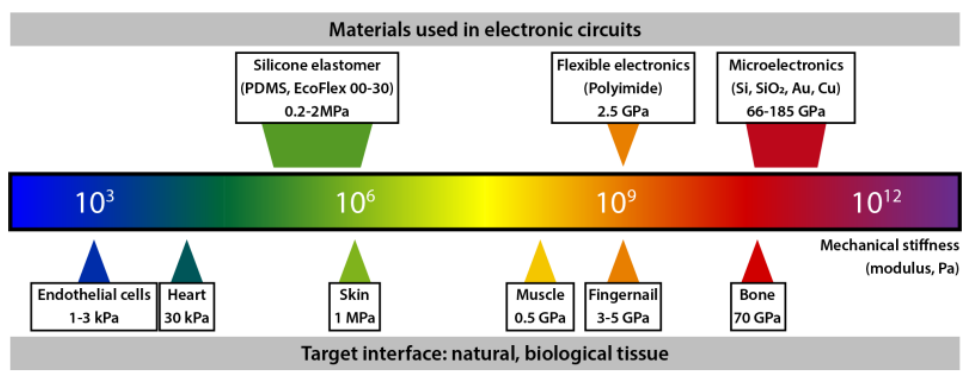

**Figure 1. Elastic modulus of biological tissues and common materials in traditional and soft electronics**—*Source: the figure adapted from [2]*.

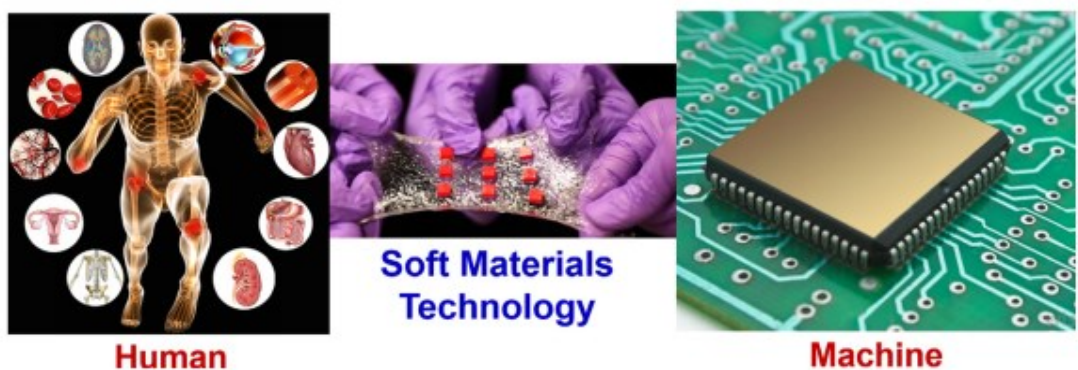

**Figure 2. Soft materials as promising materials to form a bio-compatible and safe interface between machines and human bodies**—source: the figure adapted from [3].

## 2 Soft-Matter Electronics: Background

Soft-matter electronics represent the class of electronic materials and circuits composed of soft condensed matter, which exhibit mechanical properties similar to soft biological tissue and can be easily interfaced with our body [17]. Because of this, soft-matter electronics have found various applications in the fields of bio-mimicry soft robotics, implantable devices, stretchable sensors, and artificial skin or on-skin wearable electronics for health and biopotential monitoring. **Figure 3** shows a series of examples of soft electronics applications for wearable electronics. These applications include forehead temperature mask for body temperature monitoring[18] **(Figure 3a)**, biostickers for biopotentials monitoring[19] **(Figure 3b)**, battery-free ultrathin electronic tattoo for ECG monitoring[20] **(Figure 3c)**, and LMEE-based nanotriboelectric for renewable power supply for wearable electronics [21].

However, stretchable electronics fabrication is not a straight forward process since most electrically conductive materials are rigid and soft materials are electrically and thermally insulator. To overcome these limitations, in recent decades, researchers have developed a variety of different approaches and materials architectures to create soft and stretchable electronics in which conductive materials can support large deformation. Efforts to create stretchable circuits began in the mid-1990s by introducing stretchability through geometry, called deterministic architectures [1].

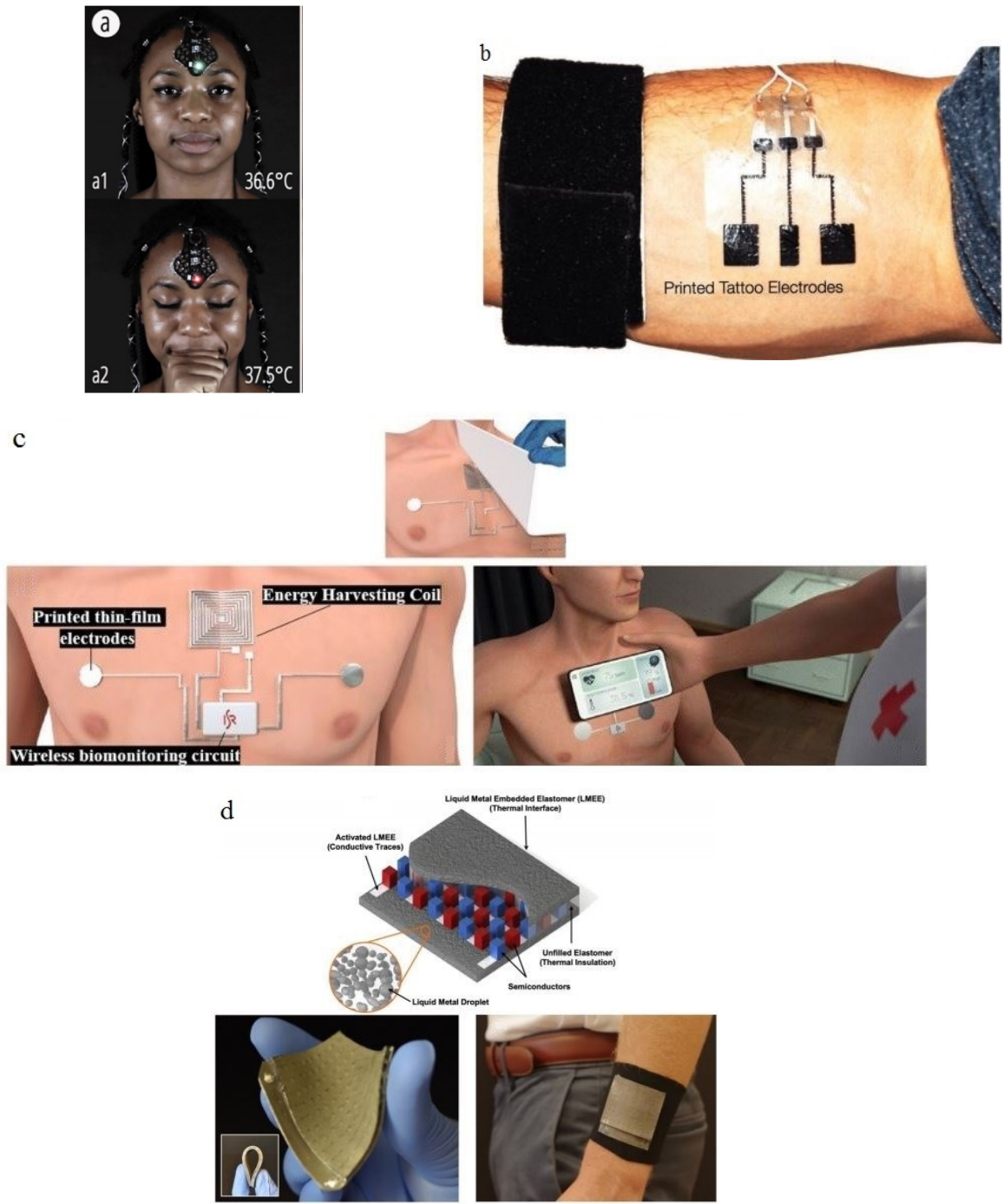

**Figure 3. Soft-Matter Wearable Electronics. a) A wearable temperature mask for real-time health monitoring (green: normal conditions and red: elevated body temperature) [18], b) an ultrathin biosticker for biopotentials monitoring [19], c) Ultrathin battery-free electronic tattoo for fully untethered wireless ECG monitoring. The biosticker receives the required energy from the mobile device, and wirelessly communicates physiological data such as heartbeat and body temperature [20]. Wearable and stretchable thermoelectric generators enabled by LMEEs for self-sustainable power supply**—Source: the figures adapted from [18-21], respectively.

## 2.1 Deterministic Structures

A high degree of stretchability along with electrical conductivity can be achieved by embedding 2D wavy or serpentine shape patterns of a thin layer of conductive metals, like gold or copper, on a soft substrate [17]. This type of stretchable electronics is known as deterministic structures or wavy circuits [17]. **Figure 4 a-b** show two examples of deterministic structures for stretchable electronics. The significant advantage of this strategy is that it can be extended into traditional microelectronics and semiconductors technology for fabricating multifunctional stretchable circuits by combining with novel lithography techniques [1]. However, since the conductive materials deposited on soft substrates are intrinsically rigid and demonstrate low elastic deformability, they may undergo plastic deformation or even rupture under large deformations, which, in turn, leads to electrical failure [17]. Notably, the mechanical properties mismatch between the soft substrate and the deposited thin metallic films can lead to stress concentration and delamination. Due to these limitations, the application of this type of conductive soft electronics is limited to low-to mid-strain operations [1, 17, 22].

## 2.2 Nanoparticles-Filled Conductive Polymers

It is possible to produce soft conductive materials by dispersing conductive nanoparticles such as carbon black, single or multiwall carbon nanotubes, graphene, $TiO_2$, Ag nanorods, and ionomers (e.g., PEDOT: PSS) within a soft matrix. In this strategy, within a soft and elastic medium, nanofillers percolate to form conductive routes. These nanocomposites combine the stretchability and compliance of soft materials with the dispersion phase's electrical and thermal properties [1, 17, 22]. In other words, while the polymeric medium enables the composite to be deformable and stretchable, the dispersing rigid nanoparticles make the composite electrically conductive. However, nanoparticles-filled conductive stretchable materials have relatively low conductivity compared to metals, which, in turn, dramatically limits their application. Although the high-volume concentration of rigid conductive particles in a soft medium may enhance the composite's electrical performance, the extreme mechanical mismatch between rigid particles and the soft matrix leads to large stress concentrations, degradation, increase in the mechanical hysteresis, decrease in the stretchability of the composite, and the inelastic behavior of the stretchable system [1, 17]. The inelasticity of the conductive composites limits the reliability and durability of the system. Additionally, in rigid particles-filled elastomers, the composite's resistivity will exponentially increase by stretching due to the separation of particles and the disintegration of the percolating network [1, 17]. **Figure 4c** exhibits a stretchable electronic system that exploits this approach.

## 2.3 Conductive Fluid-based Soft Electronics

As mentioned in the previous section, due to the high rigidity of conductive rigid particles, the resulting nanoparticle-filled composites demonstrate increased bulk rigidity and reduced stretchability [1, 17]. One alternative method to reduce this mismatch is to embed conductive ionic fluid fillers, a salt solution in water, within elastomers or gels **(Figure 4d)**. Conductive fluid-based stretchable electronics eliminate some limitations of deterministic structures and polymer nanocomposites since they eliminate the stress concentration in the interface of rigid fillers and the soft elastic medium. In this approach, the ionic solution forms continuous conductive pathways within the matrix that enables the composite to be electrically conductive because of the high fluidic volume. The composition and concentration of the salt control the conductivity and electrical properties of ionic liquids. However, these ionically conductive gels or elastomers rapidly evaporate and dehydrate without sealing with additional elastomer layers. The water evaporation causes this type of nanocomposites to dry out over time and lose their mechanical and electrical functionality. These materials also require AC electricity to prevent the fluid from electrolysis or polarization, making it difficult to integrate with standard DC electronics [2]. Furthermore, their low electrical conductivity (~1 S/m) is inadequate for most stretchable electronics applications [1, 17].

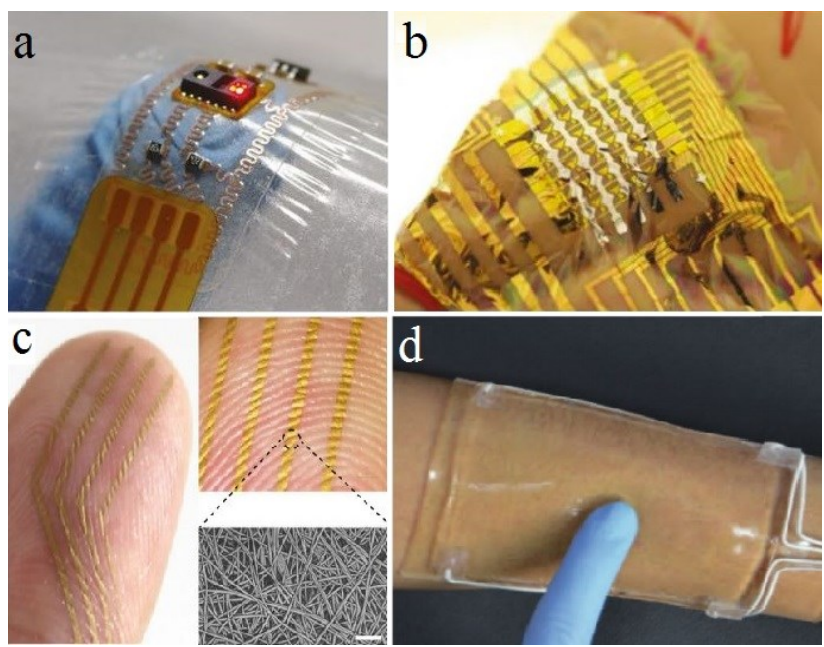

**Figure 4. On-skin stretchable electronics enabled by deterministic architectures (a-b), nanoparticles (c), and conductive fluids (d). a) nanoparticle-filled and fluid-filled composites: a) a wearable device for electrophysiological signals measuring, b) a deterministic structure enabled ultrathin organic field-effect transistor, c) on-skin conductive patterns for muscle activity measuring using conductive nanoparticles (scale bar is 5μm), d) a wearable and soft touch-panel enabled by ionic hydrogels**— *Source: the figures adapted from [23].*

## 3 Liquid Metal Embedded Elastomers (LMEEs)

Compared to rigid fillers and conductive fluids, liquid-metal droplets represent a promising alternative for the dispersion phase in a soft matrix **(Figure 5)**. since the liquid metals have a high degree of electrical and thermal conductivity, using them as fillers enhances the dielectric and thermal properties of composites [1] Embedding liquid metals in elastomers eradicates the extreme mechanical mismatch between rigid particles or deposited thin films with the soft elastic matrix or substrate. In this new class of soft composites, since the inclusion phase is liquid with low viscosity, the stress concentration at the interface of the matrix and the fillers is significantly reduced. Because of this reduction in stress concentration, the soft elastic matrix maintains its desired mechanical properties, i.e., stretchability and deformability. Importantly, this high stretchability is achieved without sacrificing the electrical or thermal properties **(Figure 6a-b)**. As shown in **Figure 6a**, while the liquid-metal composite has been stretched to 600% strain, it maintains its electrical functionality. Ga-based alloys such as a binary alloy of eutectic Ga-In (EGaIn) and ternary Ga-In-Sn (Galinstan) are popular liquid-metal alloys in LMEEs.

Kazem et al. [10] have studied the mechanical behavior of the LMEEs under tensile loading for various liquid-metal volume ratios from 0% to 50% (φ in **Figure 6**) [24]. **Figure 6c** indicates the stress-strain responses of PDMS-Galestine composites. As can be seen, the limit strain of the composites reaches as high as 600%, making LMEEs a mechanically ultrastrectable and robust conductive material. According to Figure 6, by increasing the amount of liquid metals volume ratio from 0% to 50%, the elastic modulus increases from 85 to 235 KPa. As shown in Figure 6, despite the stiffening effect of liquid-metal inclusion, the composites are similar in modulus to biological tissue [25]. Accordingly, these liquid-metal embedded elastomers (LMEE) can be designed to match the elastic and physical properties of biological tissue [1]. It should be mentioned that the functionality, mechanical, and physical properties of LMEEs depend on the topology, assembly, and interaction between liquid inclusions and the host medium [26]. Additionally, It has been proven that EGaIn-Elastomer composites that are that they exhibit relatively little mechanical hysteresis distinguish them from other nanoparticles-filled composites [1].

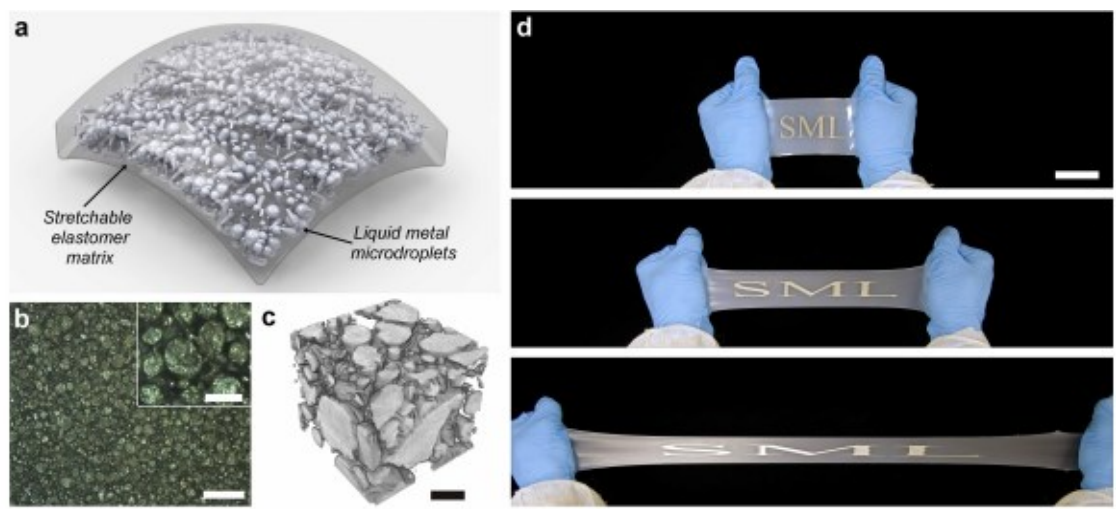

**Figure 5. Liquid Metal Embedded Elastomers (LMEEs). a) The dispersion of liquid-metal droplets within a soft elastomer, b) Optical micrograph of LMEEs with 50% volume ratio of liquid metals, c) the 3D microstructure of LMEEs, d) high stretchability of LMEEs; top: 0%, middle: 200%, and bottom: 500% strain**— Source: the figures adapted from [1]

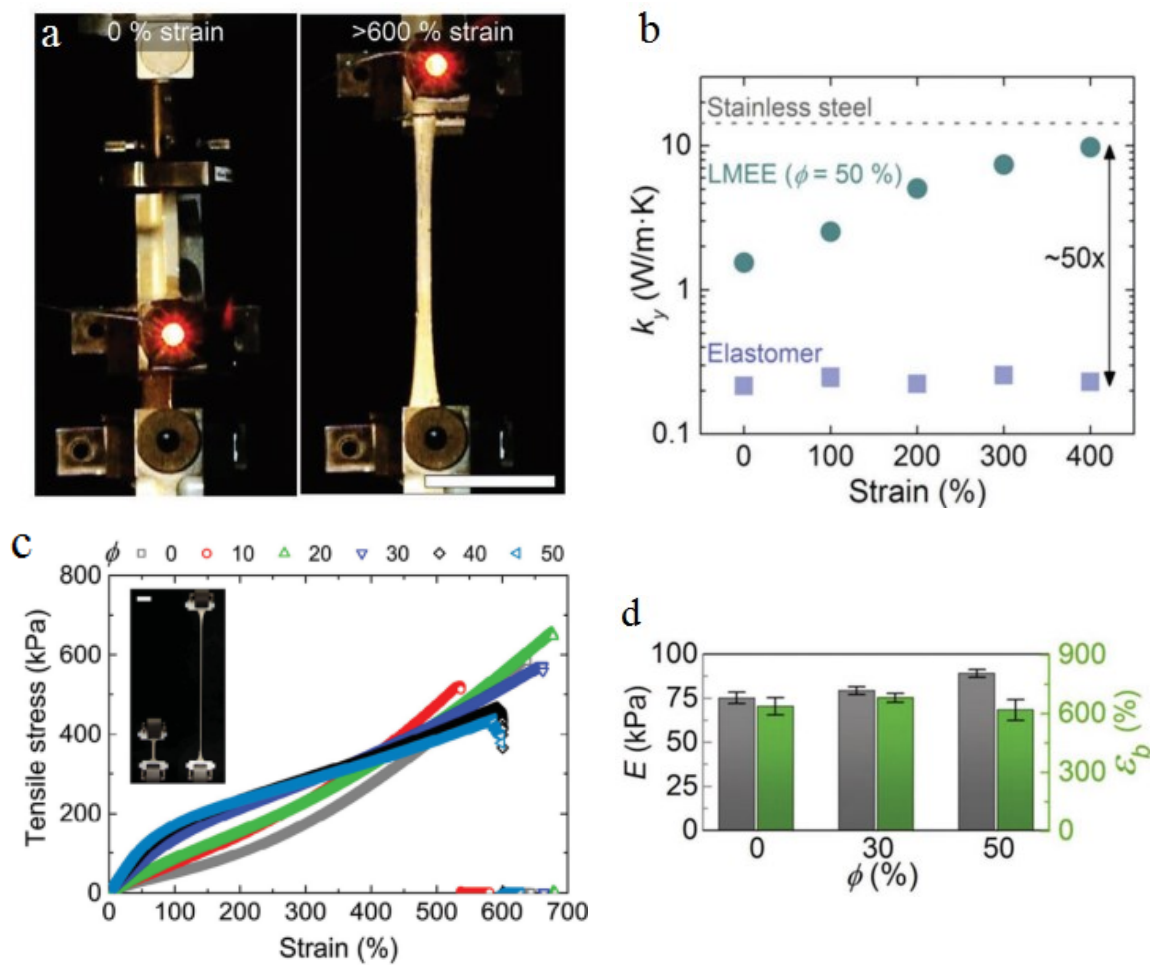

**Figure 6. Mechanical properties of LMEEs. a)** A LMEE powering an LED. While the composite exhibits ultrastretchability (>600% strain), it maintains its electrical conductivity, **b)** thermal conductivity of LMEEs with 50% volume ratio of liquid metals as a function of strain (in the direction of stretching), **c)** stress-strain response of LMEEs with the different volume ratios of liquid metal (the extension rate of the test is 10 mm/min), **d)** Elastic modulus and limit (rupture) strain as a function of liquid metals volume ratio— *Source: the figures adapted from [25].*

## 4 Materials and Preparation Process

Liquid metal embedded elastomers are composed of a soft matrix or substrate (insulating material) and liquid-metal inclusions as the conductive part of the composites.

## 4.1 Materials: Matrix Phase

Elastomers and polymers are typically used as the matrix (or substrate) material for liquid-metal embedded multifunctional composites. Common matrix or carrier materials include silicone elastomers, in particular polydimethylsiloxane (PDMS), an easy-to-process organic silicone which is fully transparent, polyurethanes (PU), soft polyacrylates, elastomeric fluoropolymers, and styrene ethylene butylene styrene copolymer (SEBS) [27]. Pt-cured Ecoflex 30 and Sylgard 180 are of particular interest PDMS elastomers in stretchable electronics [14]. It should be mentioned that depending on the microstructural characteristic of elastomers, such as polymer chain lengths and cross-linking density, stretchability differs among elastomers [22]. Accordingly, the elastic strain limit (rupture strain) of elastomers can range from 100% to as high as 900% [22]. In addition to high deformability and stretchability, another critical feature of elastomers such as PDMS is that these materials are nontoxic and biocompatible.

## 4.2 Materials: Dispersion Phase

In LMEEs, liquid metals are the dispersion phase of the composite material. Liquid-metal alloys possess the unique combination of the low viscosity of fluids and the electrical and thermal properties of metals [17]. The low viscosity of liquid-phase metals make them compliant to deformation and injectable to soft microfluidic channels, which are of great interest features in stretchable electronics.

In comparison to other liquid metals like Hg, Fr, and Cs, Ga-based alloys such as eutectic Ga-In (EGaIn) and Ga-In-Sn (Galinstan) has gained more attraction because of their low melting point, stability in room temperature **(Figure 7a)**, low toxicity (compared to Mercury), low viscosity (2mPa·s), high electrical and thermal conductivity, and negligible vapor pressure [10]. Furthermore, in the presence of oxygen such as air or oxygen-saturated elastomers, Ga-based liquid metals form a self-passivating $Ga_2O_3$ oxide layer (with the thickness of ~0.5-3 nm), an n-type semiconductor with a bandgap of approximately 4.85eV. This oxide layer dramatically reduces the surface tension of droplets and acts as an elastic membrane until the surface stress exceeds a critical value of 0.5–0.6 $N/m^{-1}$[28]. When the layer breaks, it reforms instantaneously, allowing droplets to be structurally self-stabilizing, printable, and moldable **(Figure 7b)**. Additionally, this oxide skin increases the adhesion between GaIn and the substrate and will enable droplets to hold their shape [29]. The microdroplets of liquid metals are generally ellipsoidal-shaped with dimensions on the order of ∼ 4-15 $\mu$m [1]. The physical properties of these alloys are presented in Table. 1.

**Table 1. Physical properties of common liquid metals for stretchable electronics [1, 17]**

| Liquid Metal | Composition (wt %) | Conductivity | Melting Temperature | Thermal conductivity | Surface Tension |
|---|---|---|---|---|---|
| Eutectic Gallium-Indium | 75.5% Ga 24.5% In | $3.46 * 10^6$ S/m | 15.5 C | 26.4 W/m.K | 624 |
| Gallium-Indium-Tin (Galinstan) | 68.5% Ga 21.5% In 10% Sn | $3.40 * 10^6$ S/m | -19 C | 25.4 W/m.K | 534 |

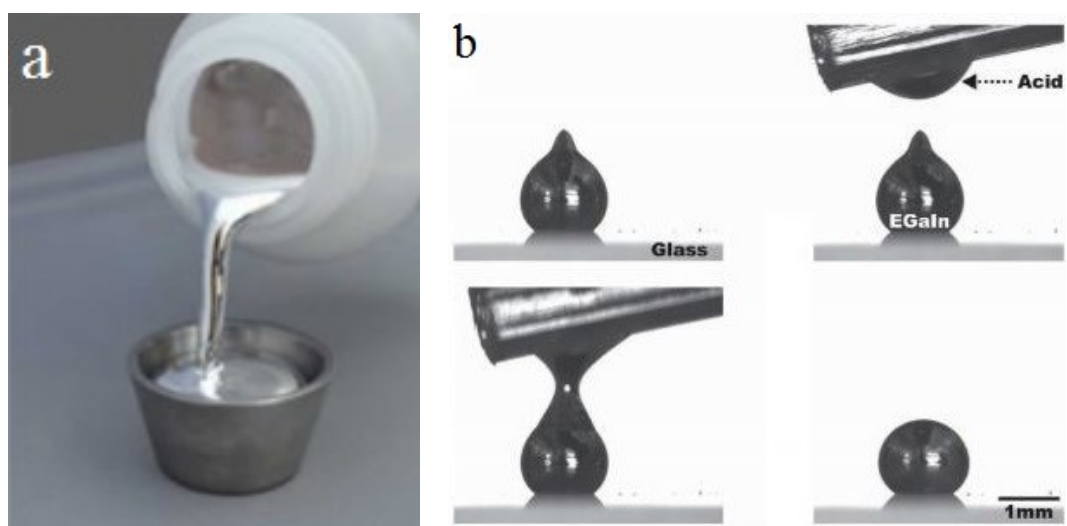

**Figure 7. a) Eutectic Gallium-Indium (EGaIn) alloy at room temperature, b) (left) the formation an oxide layer makes the liquid-metal droplets moldable and printable, (right) acid removes the oxide layer—** *Source: the figures adapted from [1, 30], respectively.*

### 4.3 Preparation Process of LMEEs

Producing a dispersion of liquid metal droplets within a solvent is the first step to prepare LMEEs. Various methods have been developed to produce liquid metal droplets from mm to nm scale [8]. **Figure 8** illustrates a series of common approaches to produce liquid-metal droplets. These common synthesizing methods include molding (**Figure 8a**), using acoustic waves (**Figure 8b**), employing microfluidic devices (**Figure 8c**), shear mixing (**Figure 8d**), and ultrasonication (**Figure 8e**).

After synthesizing liquid-metal microdroplets, LMEEs are prepared by mixing the droplets with an uncured elastomer (e.g., PDMS) at the desired volume ratio of liquid metals (e.g., 50%). The mixing is continued until the formation of a viscous emulsion with a dimension of microdroplets of less than approximately 20-30 $\mu$m [1]. **Figure 9** indicates the effect of mixing time on droplet size. Increasing the mixing time decreases droplet size. Despite a significant difference between the densities of inclusions and the host matrix, the Ga-based alloys exhibit uniform dispersion. They remain suspended within the PDMS because of their relatively high concentration [9]. **Figure 10** indicates an almost even distribution of microdroplets and the absence of percolating conductive networks. The presence of percolating networks could result in electrical conductivity or shorting.

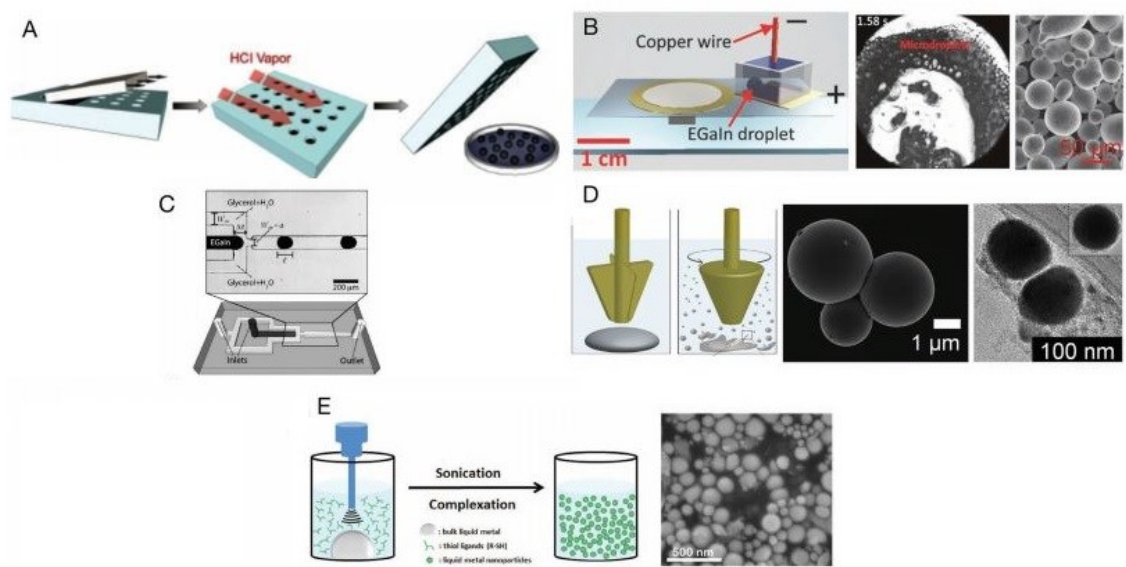

**Figure 8. Common methods to synthesize the droplets of liquid metal: a) Molding, b) acoustic wave-induced forces can be used to produce microdroplets, c) liquid metal synthesis using microfluidics, d) shear mixing, e) ultrasonication**— *Source: the figures adapted from* [10]

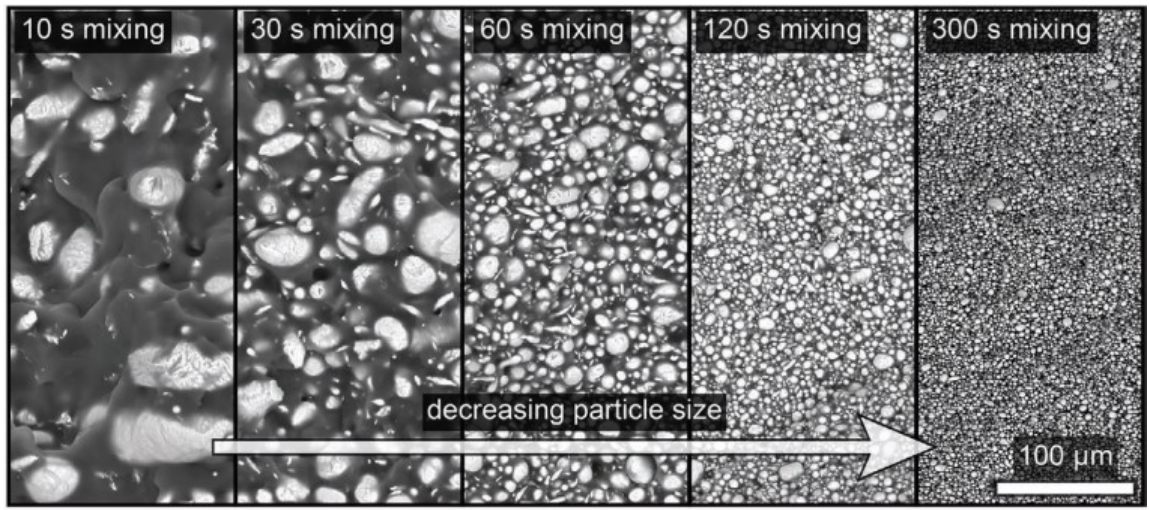

**Figure 9. Cross-sectional images of liquid-metal embedded PDMS (Sylgard 184). Increasing liquid metal-uncured PDMS mixing time leads to a decrease in droplets size** *—Source: the figure adapted from [31]*

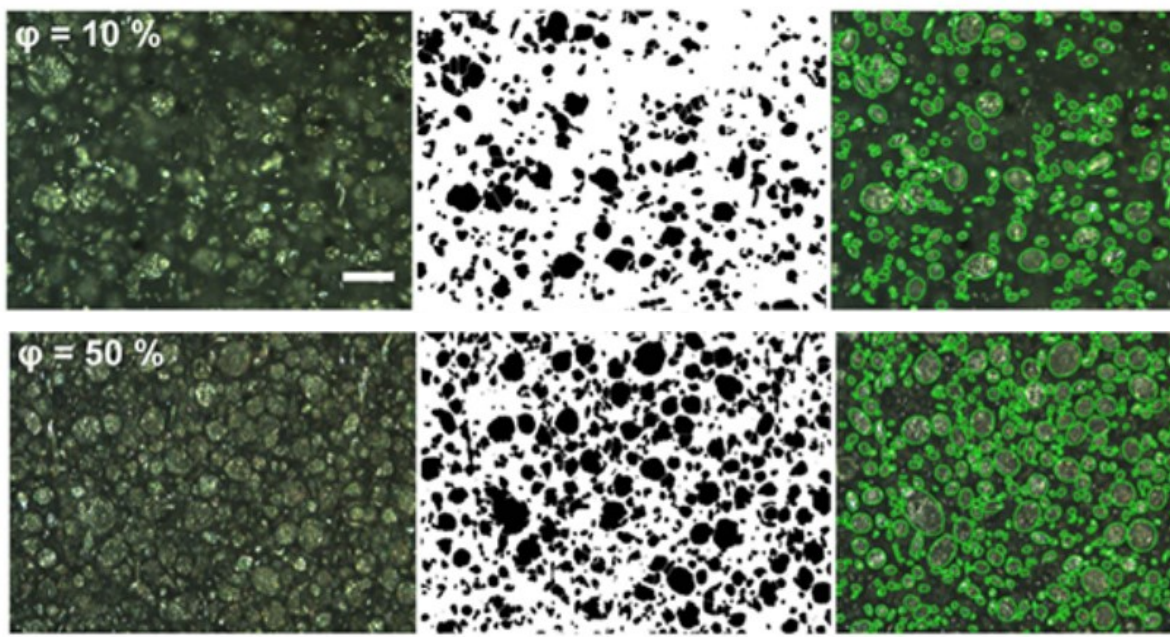

**Figure 10. Optical micrographs of LMEEs with the liquid-metal volume ratio of $\varphi = 10\%$ (top) and 50% (bottom). The micrographs indicate an almost uniform distribution of liquid-metal droplets within the soft medium. The scale bar is 50 $\mu$m, and the size of the analyzed area is 400 × 300 $\mu$m**. *—Source: the figure adapted from [1].*

## 4.4 Activation Process of LMEEs

After initial mixing and preparation of LMEEs, the initial composite is non-conductive because of an insulating oxide layer formed on the liquid metals' top surface and the non-coalesced dispersion of microdroplets in the insulator matrix. For LMEEs to demonstrate volumetric conductivity, liquid-metal inclusions must coalesce within the soft medium to form a continuous liquid metal pathway. This can be achieved by applying concentrated mechanical pressure, known as mechanical sintering **(Figure 11)**. The localized pressure results in the rupture of liquid metals and the soft medium's internal tearing, which, in turn, leads to the coalescence of droplets and the formation of permanent conductive pathways. The process of creating conductive pathways is known as trace pattering. These traces can be used for power and data transmission within the composite. The trace pattering process allows inducing conductivity in the material selectively and patterning multiple traces that are electrically insulated from each other.

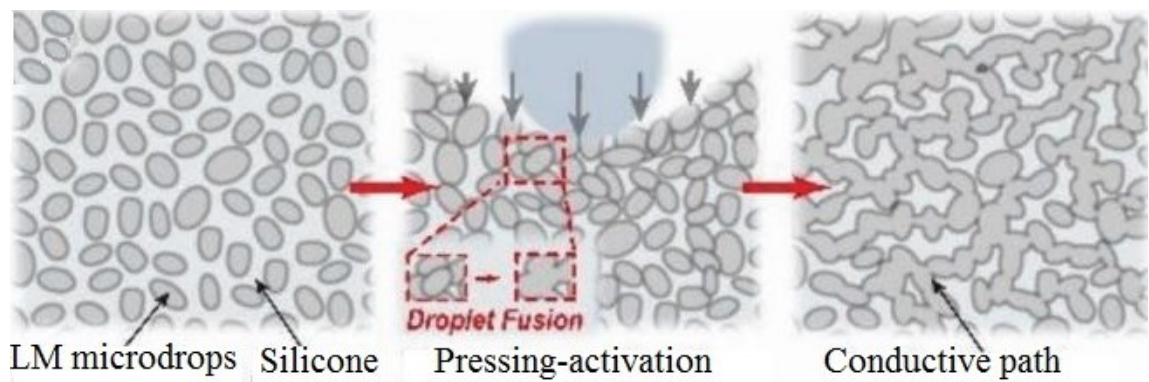

**Figure 11. Mechanical activation of LMEEs to create conductive pathways within a silicone matrix**—*source: the figure adapted from [32]*

## 5 Patterning Methods of Liquid-Metal Electronic Circuits

In the presence of oxygen, liquid-metal alloys form a self-passivating thin oxide layer of $Ga_2O_3$. The oxide layer allows patterned features to hold their shape. Beyond the critical surface tension, the oxide layer breaks, and the liquid flows readily. High moldability and ability to adhere to various surfaces have enabled Ga-based alloys to be patterned with multiple low-cost, room temperature, ambient pressures, clean-room free, and fast fabrication techniques that are compatible with various types of substrates without the need for vacuum processing (e.g., PVD or CVD techniques) [33].

## 5.1 Injection Filling

Injection filling is one of the most common techniques for producing liquid metal circuits. **Figure. 12** shows a fabrication process for producing a circuit by injection filling. According to Fassler et al. [17], in the injection filling process, first, a patterned elastomeric substrate **(Figure 12c)** is fabricated by casting the uncured material **(Figure 12b)** in a mold with the desired microchannels geometry of the substrate **(Figure 12a)**. Common methods to create the mold are 3D printing (additive techniques), photolithography, or machining processes (subtractive processes) such as traditional milling or laser machining. Next, the microchannels are sealed by bonding the pre-patterned elastomer to an additional layer of an elastomer **(Figure 12d)** [17]. After sealing, a needle and syringe are used to inject the liquid metal into the inlet holes of the pre-patterned microchannel microfluidic channels **(Figure 12e)**. After injection, the oxide layer adheres to the channel surfaces, which, in turn, results in a stabilized microstructure. Finally, external wires inserted into the circuit terminals to complete the process **(Figure 12f)**.

Since the microchannels initially contain oxygen, the geometry of channels must be continuous with both an inlet and an outlet to give vent to air as the liquid metal is inserted. While the method is reliable for producing simple circuits with a single inlet and outlet, injection filling cannot be extended to fabricate intricate patterns such as self-intersecting or close-loop patterns. Also, creating a mold using SU-8 photolithography or rapid prototyping tools is time-consuming [34].

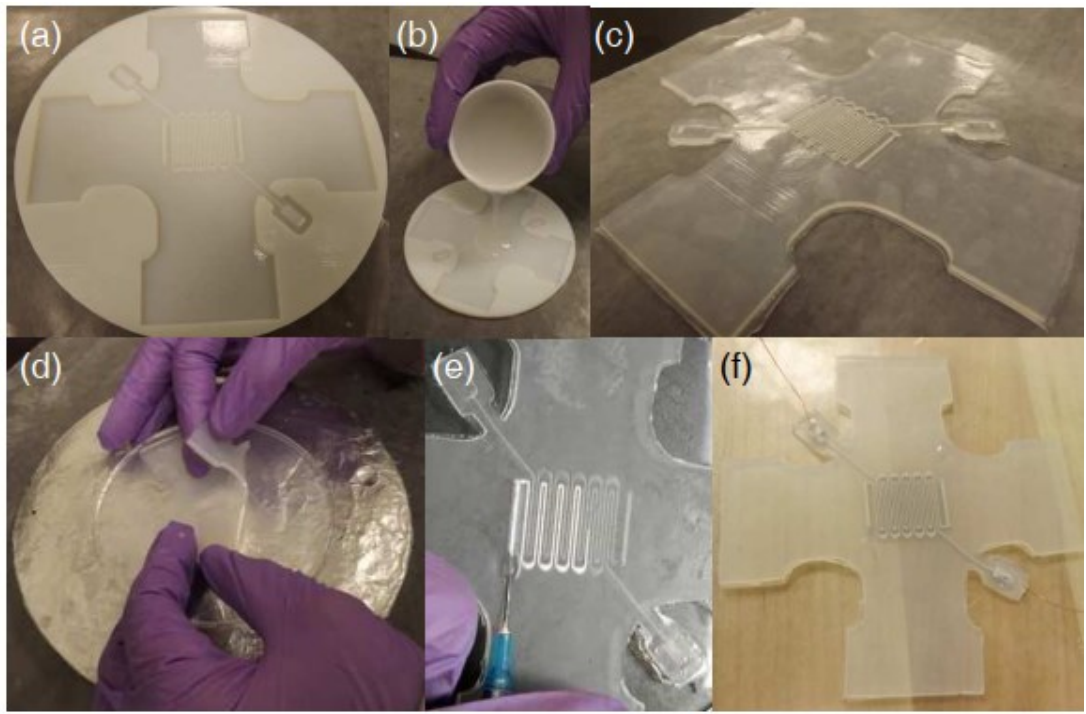

**Figure 12. Injection Filling. a) a 3D printed mold, b) the mold is filled with uncured elastomer, c) the elastomer is cured and released from the mold, d) the released elastomer is sealed, e) GaIn is injected into microchannels, f) external copper wires are inserted—** *Source: the figures adapted from* [17]**.**

## 5.2 Stencil Lithography

Stencil lithography is a simple process that can be performed in clean-room free conditions **(Figure 13)**. In this lithography-enabled process, liquid metals are deposited on an elastomer substrate using a patterned stencil mask. In summary, a mask (stencil) with pre-defined patterns is applied to an elastomer substrate **(Figure 13a),** and then the liquid metal is deposited over the mask **(Figure 13b)**. The liquid metal deposits on the exposed parts of the substrate. After removing the mask, the deposited liquid metals remain, and the desired pattern is obtained **(Figure 13c)**. While the process is quite simple, it is limited by the types of planar geometries that can be patterned on the mask. For instance, the process cannot produce self-intersecting patterns or other closed-loop features [22]. Moreover, each new circuit design requires its own mask, and fabrication processes for producing masks are time-consuming.

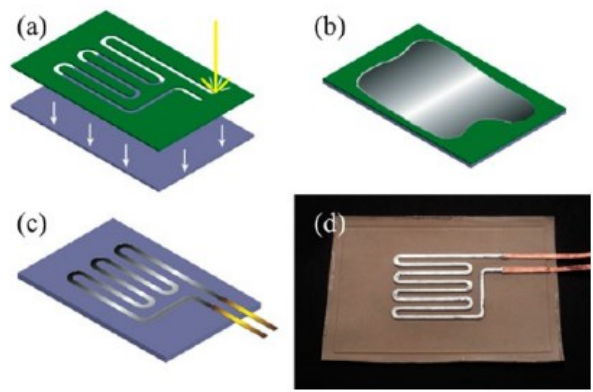

**Figure 13. Stencil Lithography. a) a laser machined mask (stencil) is placed on a soft substrate, b) liquid metal is deposited on the surface, c) after removing the mask, copper wires are inserted, d) a GaIn heater on VHB elastomer fabricated by stencil lithography—** *Source: the figure adapted from* [34]**.**

## 5.3 Imprinting

Imprinting with soft molds is another lithography-enabled simple patterning technique. In summary, a thin layer of liquid metal is deposited on a flat surface, and then a soft mold (e.g., PDMS) with topographical raised patterns is pressed against the liquid metal layer **(Figure 14a)**. Consequently, with the help of the oxide layer, a thin film of liquid metal adheres to the raised features of the soft mold **(Figure 14b)**. Therefore, after removing the mold form the substrate, the liquid metal remains on the topographical features of the mold. Finally, the circuit is seald with the help of another an elastomer layer **(Figure 14c)**.

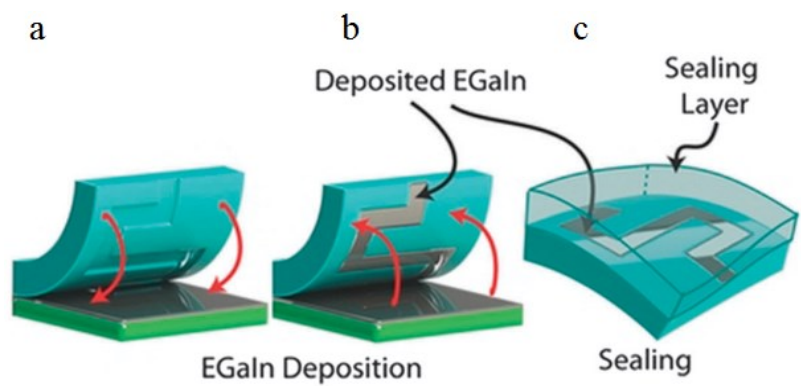

**Figure 14. Imprinting. an elastomeric mold is pressed against a thin layer of liquid metal, and therefore, liquid metal adheres to the features of patterned mold—** *Source:the figure adapted from [28].*

## 5.4 Microcontact Printing

Microcontact printing is an inexpensive and clean-room free technique for depositing soft and conductive inks on a soft substrate. There are two primary methods to pattern liquid metals using micro-contact printing: 1- Stamp printing or entire transferring of ink using a topographical stamp, and 2- successive deposition of individual droplets of liquid metal with a hemispherical print head. In contrast to stencil lithography and injecting filling, microcontact printing (stamp printing and successive deposition) can be used to create circuits with any planar pattern with intersecting and closed-loop circuit features [28, 34]. The process benefits from the oxide layer, which preserves the shape of the deposited droplets of liquid metal.

### 5.4.1 Microcontact Printing - Stamp Printing

In the first method of microcontact printing, i.e., stamp printing, a PDMS stamp with pre-patterned features of the desired geometry is pressed against a pool of GaIn **(Figure 15a; 1-4).** The liquid metal adheres only to the raised features of the mold. Then, the stamp is pressed against the target soft substrate **(Figure 15a; 5)**, which, in turn, results in the entire transferring of the liquid metal into the soft substrate **(Figure 15a; 6)**. Eventually, an elastomer layer is bonded to the PDMS substrate to seal the circuit **(Figure 15a; 7-8)** [35]**.**

### 5.4.2 5.4.2. Microcontact Printing – Successive Deposition of Droplets

In the second microcontact printing method, circuits are printed droplet by droplet rather than entire pattern transferring. In this method, which we call "successive deposition of droplets," a soft needle with a hemispherical tip dips into a liquid metal pool. Consequently, a layer of liquid metal is formed on the print head. The needle tip then touches the substrate to deposit the droplets of liquid metal in a pre-defined path **(Figure 15b)**. Subsequently, successive deposited liquid-metal droplets merge to form a patterned circuit on the elastomer substrate. After printing circuits and external wiring insertion, the patterned elastomer is sealed with an additional layer of elastomer. The microcontact printing process can be automated by implementing the print head on a motorized 3-axis platform [34].

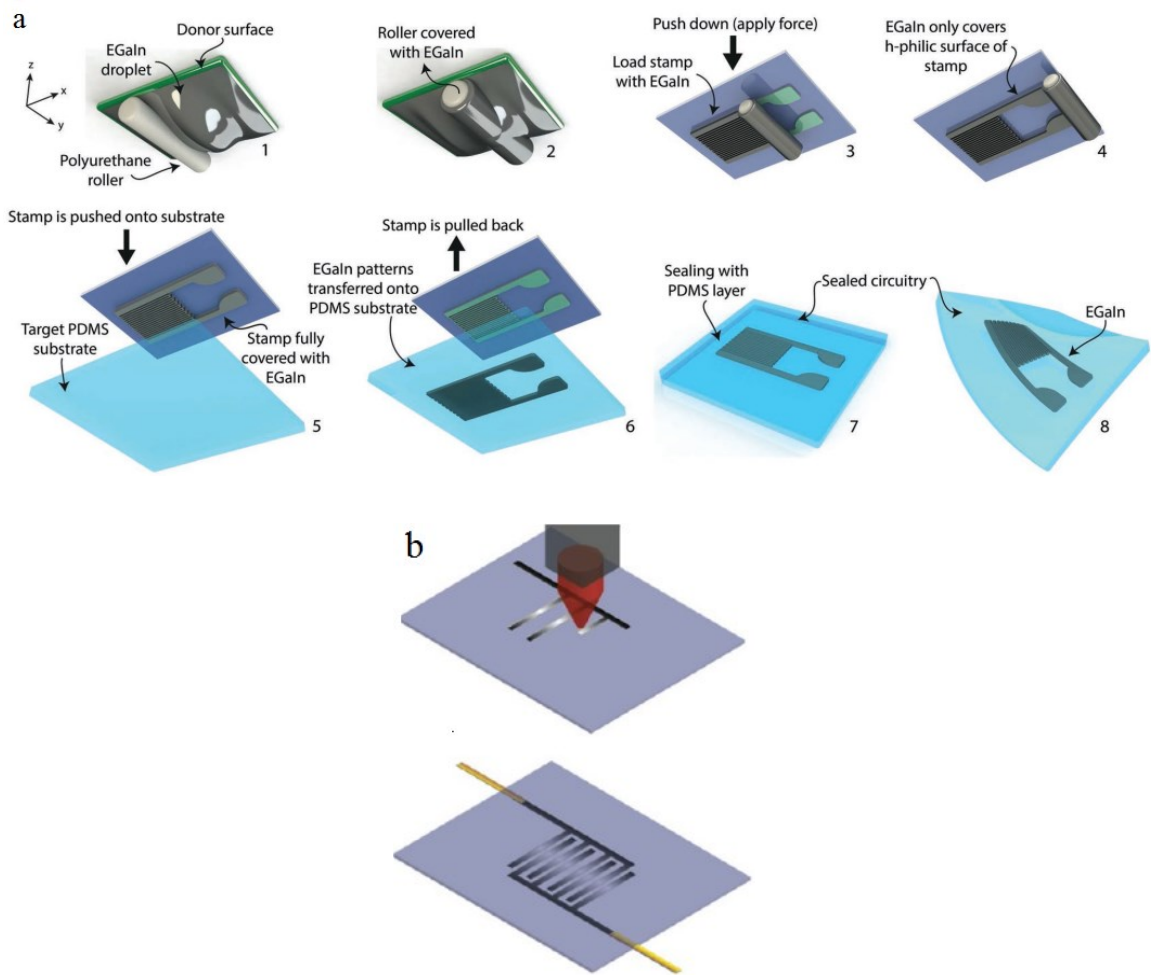

**Figure 15. Two methods of microcontact printing. a) Stamp printing, b) Successive deposition of droplets**— *Source: the figures adapted from [34, 35], respectively.*

## 5.5 Direct-Write Printing (Extrusion)

Direct-write printing is an extrusion-based additive patterning technique that can produce 2D and 3D patterns with heights up to ~1 cm in a completely automated and rapid manner **(Figure 16)** [30]. In this process, liquid metal, in the form of droplets and under pressure, is extruded onto a substrate through a nozzle. The formation of the oxide layer bring about a stabilized structure. The inner diameter of nozzle, distance between the nozzle and substrate, and the flow rate of the extruded material are important parameters of the direct-writing process, which affect the quality and geometry of traces [28, 34]. Since direct-write printing does not involve lithography steps for mold fabrication, it is a fast pattering method which is able to pattern any pattern in a few steps.

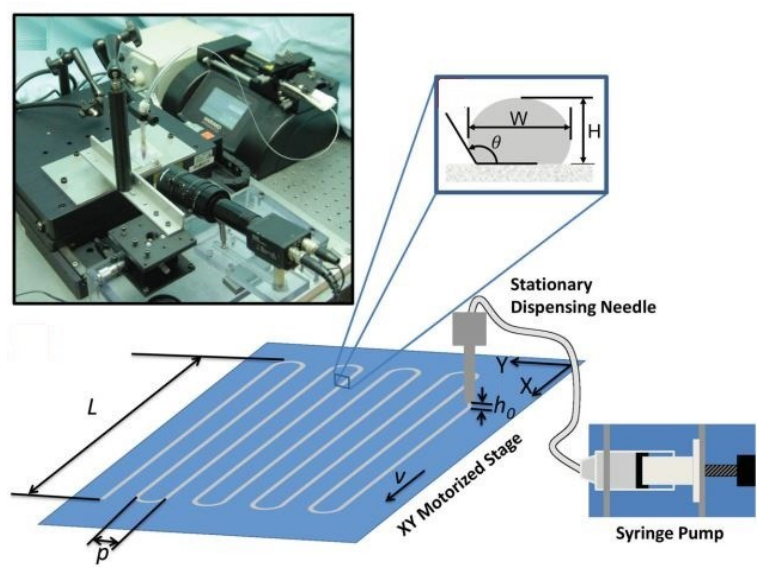

**Figure 16. Direct-Write Printing**— *Source: the figure adapted from [30].*

## 5.6 Laser-based Patterning

$CO_2$ or UV lasers have been widely employed in various manufacturing and processing applications such as machining, macro- or nano-scale welding, tissue surgery, microfluidic channels fabrication in polymers, and mask fabrication lithography applications, and thin-film patterning in stretchable electronics [22]. Because of its high precision in the pattering of materials, lasers have become a popular and versatile tool in pattering liquid metals in soft electronics [22, 27, 36, 37].

Liquid-metal pattering by laser is a subtractive patterning technique and is initiated by sealing the liquid metal layer between two elastomer sheets (e.g., PDMS) **(Figure 17a)**. The sealing prevents the liquid metal from oxidation and protects the top surface of the composite from debris and residuals of the process after vaporization and explosion [22]. In order to selectively remove the liquid metal, a $CO_2$ laser (wavelength =10.6 μm) scans over the entire surface and leads to local heating **(Figure 17b)**. Materials absorb the photonic energy of the $CO_2$ laser, which, in turn, results in the PDMS layers evaporation **(Figure 17c)**. When the pressure difference between the vaporized PDMS and the atmosphere exceeds the surface tension of the oxide layer of the liquid metal, the escaping vapor puncture the liquid metal layer and displaces it away from undesired regions **(Figure 17d)**.. In contrast to elastomers, liquid metal alloys cannot be ablated with a $CO_2$ laser since they require a high-energy UV laser for ablation [22]. The oxide layer, i.e., $Ga_2O_3$, prevents the liquid metal from flowing back into the laser-machined regions.

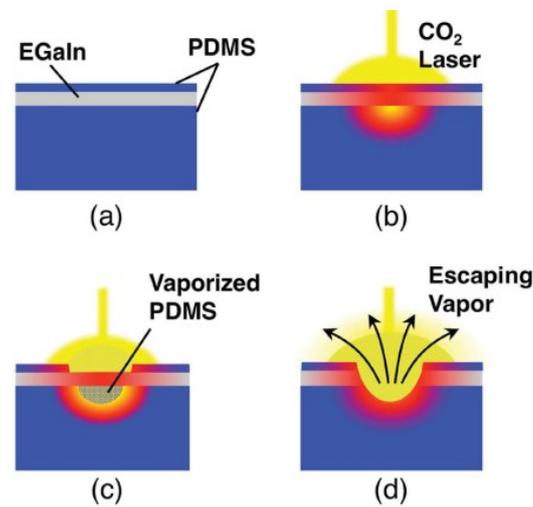

**Figure 17. The laser-based pattering of liquid metals. a) A layer of PDMS is coated on liquid metal to prevent oxidation and protect the top surface from the process's residuals. b) Local heating of the sample during laser scanning, c) PDMS vaporization d) The escaping vapor of PDMS punctures the liquid metal**— *Source: the figure adapted from [22].*

## 6 Case Study: Autonomously Self-healing Circuits

As it occurs in real-world applications, soft robots and electronics are subjected to various extreme mechanical damaging such as rupturing, tearing, and puncturing that may result in electrical failure. The electrical failure of soft electronics due to mechanical damaging may limit the applications of these types of materials in fields like autonomous robotics [38]. Thus, it is crucial to develop a new class of soft electronic materials that are autonomously self-healing and remain functional even as the material is mechanically damaged [38, 39].

Using LMEEs and mechanical sintering, Markvicka et al. [38] have developed a new design for liquid-metal embedded elastomers to produce electrically self-healing soft circuits that preserve their electrical functionality while immensely damaging under various mechanical loadings **(Figure 18)**. When damage occurs, the dispersed GaIn microdroplets in the soft medium rupture to coalescence with neighboring microdroplets **(Figure 18b)**. Damage induced new droplet–droplet connections create new pathways and reroute the electrical conductivity around the damaged area **(Figure 18b)** [38]. The main advantage of this type of self-healing composites is that electrical conductivity is rerouted spontaneously without any need for external interruption (e.g., manual reassembly or external heating). This unique self-healing property is shown in **Figure 18c**, which shows that while the power, data, and clock lines of a four-channel serial clock display immensely damage by cutting, tearing, and the complete removal of material, it maintains its operation [38].

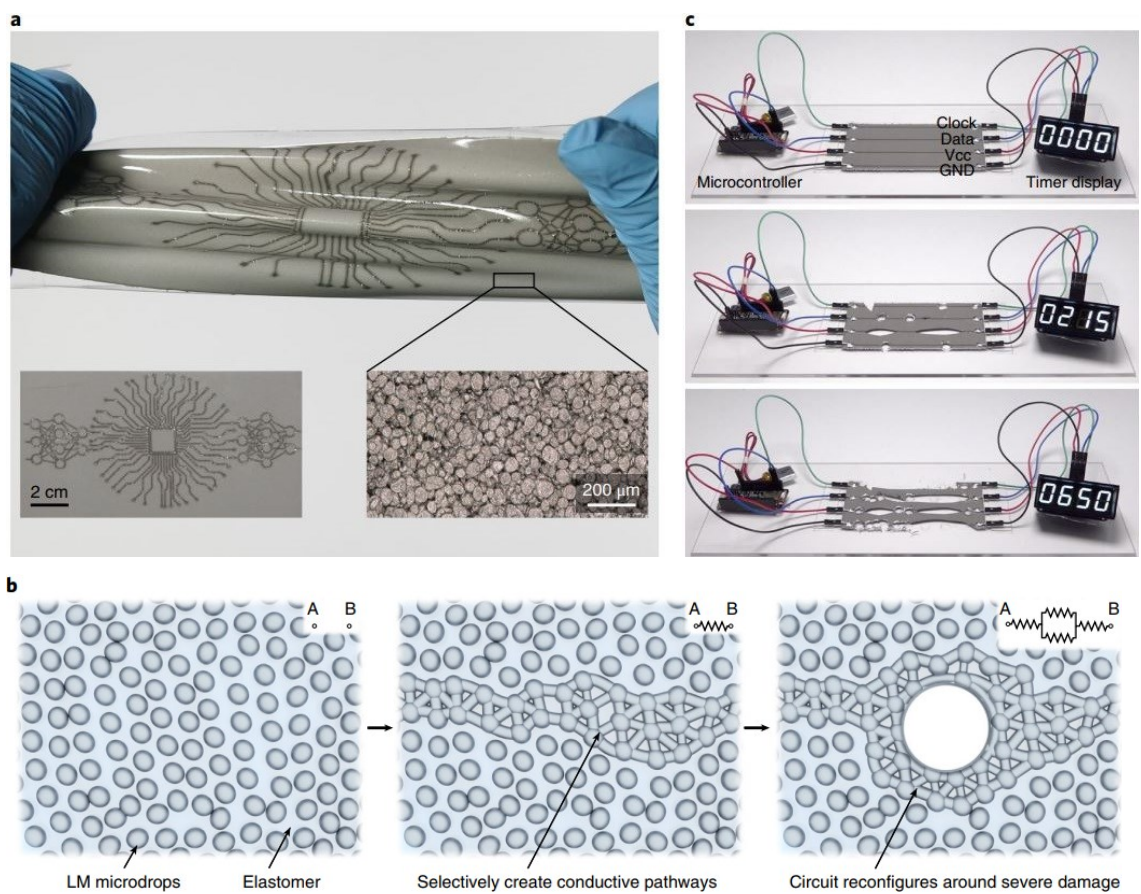

**Figure 18. Autonomously self-healing liquid-metal embedded elastomers. a) A stretchable and deformable self-healing liquid-metal circuit with electrically conductive traces, b) The self-healing mechanism; mechanical damage results in the autonomous formation of new conductive pathways around the damage. c) Self-healing LMEE-enabled system transmitting DC power. While the conductive traces is extremely damaged, the systems maintain its operation**— *Source: the figure adapted from [38]*.

## 6.1 Fabrication Process of Self-healing LMEE Circuits

According to Markvicka et al. [38], in order to fabricate a seal-healing LMME system, first, PDMS is prepared at a 5:1 oligomer to curing agent ratio using a centrifugal mixer. Mixing and defoaming each takes 1 min. To produce EGaIn, Gallium and indium are combined at 75% and 25% (by weight), respectively. To fabricate the LMEE, PDMS and EGaIn are combined at the desired liquid-metal volume ratio, such as 50%, which corresponds to a liquid metal to elastomer mass fraction of 6.44:1. The PDMS prepolymer and EGaIn are mixed until the formation of an emulsion with small and microscale droplets of EGaIn. The emulsion is further mixed (mixing time: 1 min) to produce a polydisperse suspension of EGaIn microdroplets dispersed in the PDMS matrix. After mixing, a 300μm layer of PDMS is deposited on a polyethylene substrate and cured at 100°C for 30 min. After allowing the substrate to cool, a 550μm layer of LMEE is cast on the top surface of the PDMS using the stencil lithography patterning technique. Next, the stencil mask is removed, and the layer is cured at 100 °C for one hour. After LMME deposition, the LMEE is selectively activated using a 2D plotter **(Figure 19)** to form conductive pathways by applying concentrated mechanical loading. Trace pattering is followed by cleaning the residual liquid-metal microdroplets on the surface of the circuit using isopropyl alcohol. Finally, the circuit is sealed by an additional elastomer layer to prevent further trace formation [38].

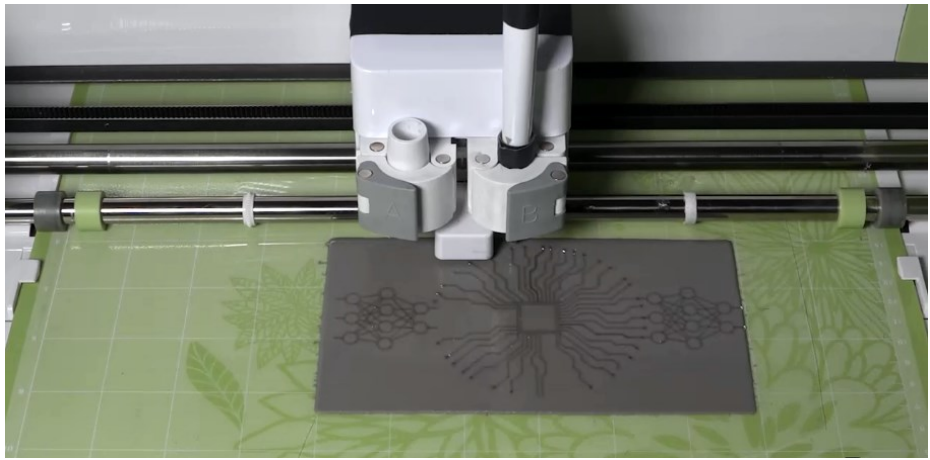

**Figure 19. A 2D plotter for trace pattering on self-healing LMEEs**— *Source: the figure adapted from [38].*

**Conclusions**

This manuscript reviewed the fundamental concepts in Liquid-Metal Embed Elastomers (LMEEs), which are highly on-demand in emerging fields like wearable devices and soft robotics. We provided a brief background of various developed methods to produce soft conductive materials. We also discussed the advantages of LMEEs over other methods. These advantages include eliminating stress concentration and degradation at the filler-matrix interface, ultrastretchability (over 600% strain) without sacrificing electrical and thermal conductivity, metal-like electrical and thermal properties, vacuum- and cleanroom-free inexpensive processing techniques, and the self-healing capability. The autonomous droplet-droplet connection within the soft medium gives LMMEs self-healing property, a critical feature for further progress in various emerging files such as autonomous soft robotics. As mentioned in the main text, the formation of the oxide layer of $Ga_2O_3$ makes the Ga-based liquid-metal droplets printable and moldable. Specifically, The presence of the oxide layer helps microdroplets maintain their shape after pattering on soft substrates. We reviewed several common micropatterning techniques of liquid metals, including injection-filling, stencil lithography, imprinting, microcontact printing, direct writing, and laser-based pattering. The merits and demerits of each method were explained separately. Despite tremendous research that has been done in the LMEEs domain, there are several yet-to-be-addressed issues, such as challenges associated with the mass-production of liquid metal-enabled soft electronics or their reliability in real-world applications. We hope this short review provides a good background for further progress and successes with LMEEs.